\documentstyle[aps,prb,floats,twocolumn]{revtex}
\input{epsf}	
\begin{document}
\draft
\wideabs{
\author{Y. Lin, J. Sichelschmidt, J. E. Eldridge}
\address{Department of Physics and Astronomy,
University of British Columbia, Vancouver, B.C. V6T 1Z1, Canada}
\author{T. Wahlbrink}
\address{Fachbereich Physik der Universit\"{a}t Osnabr\"{u}ck, 49069 Osnabr\"{u}ck, Germany}
\author{S.-W. Cheong}
\address{Department of physics, Rutgers University, Piscataway, New Jersey 08855\\
Bell Laboratories, Lucent Technologies, Murray Hill, New Jersey 07974}
\title{Observation of magnetic order in La$_{1.9}$Sr$_{0.1}$CuO$_4$ from two-magnon Raman scattering}
\date{\today}
\maketitle
\begin{abstract}
We report two-magnon Raman scattering from La$_{1.9}$Sr$_{0.1}$CuO$_4$, which has a suppressed T$_c$=12 K, as 
the temperature is lowered below 37 K and an ordered spin phase is formed. The two-magnon
Raman intensity increases with decreasing temperature. The magnetic scattering in La$_{1.9}$Sr$_{0.1}$CuO$_{4}$
is totally different from that reported in the parent compound La$_2$CuO$_4$. We analyze the line shape of 
the two-magnon scattering within the traditional Loudon-Fleury theory and find the superexchange 
constant {\bf J}=1052 cm$^{-1}$. The calculation of the frequency moment suggests that the 
quantum fluctuations are very weak in the system. The room temperature Raman scattering from La$_2$CuO$_4$ 
is also measured. Strong features appear in the one-phonon spectrum at the frequencies  of the longitudinal
optical (LO) infrared modes which we suggest become Raman active through a Fr\"{o}hlich-interaction.  
\end{abstract}
}
\narrowtext
\section{Introduction}
One of the most important questions in the study of the cuprate superconductors
is the understanding of the phase diagram of these materials,
especially the interplay between magnetism and superconductivity (SC).
La$_{2}$CuO$_{4}$, which is the parent compound of the ``214" high-T$_{c}$ cuprate
superconductors La$_{2-x}$Sr$_{x}$CuO$_{4}$ and La$_{2-x}$Ba$_{x}$CuO$_{4}$ with the simple
structure of single CuO$_{2}$ layers, is an insulating antiferromagnet with
a transition temperature T$_{N}$ around 300 K and spin {\bf s}=1/2, localized
on Cu atoms in the CuO$_{2}$ planes. With increasing dopant x above 0.06,
La$_{2-x}$Sr(Ba)$_{x}$CuO$_{4}$ become superconducting. However, at x around 0.11
in La$_{2-x}$Sr$_{x}$CuO$_{4}$ and x around 0.125 in
La$_{2-x}$Ba$_{x}$CuO$_{4}$, the superconductivity is suppressed and
magnetic ordering reemerges.\cite{mood1,axe,luke1,wat,luke2,kuma,oh,go1,go2,suzuki}\\

In the La$_{2-x}$Ba$_{x}$CuO$_{4}$ (x around 0.12), a structural
phase transition, from the low temperature orthorhombic phase (LTO) to
the low temperature tetragonal phase (LTT), has been observed
at T=70 K.\cite{axe,ko}
Below 36 K,  the magnetic ordering was observed by $\mu$SR and NMR
measurements.\cite{luke1,kuma,go2}
In the La$_{2-x}$Sr$_{x}$CuO$_{4}$ (x=0.12),  Moodenbaugh {\em et al}\cite{mood2}
observed an incomplete phase transformation (approximately 10 $\%$) below 80 K
from the LTO phase to the LTT phase, using
high-resolution synchrotron x-ray powder diffraction. The LTT phase
was also indicated by electron diffraction and ultrasonic
measurements.\cite{suzuki,ko,fuk}  There have been several
reports of magnetic ordering in La$_{2-x}$Sr$_{x}$CuO$_{4}$ (x around 0.11) at
low temperatures.\cite{wat,kuma,oh,go1,go2,suzuki}. The $^{39}$La and $^{63/65}$Cu-NMR
showed an antiferromagnetic ordering below 32 K\cite{go2} and
the Cu-spin moments were found to be in the CuO$_{2}$ plane perpendicular to
the c-axis.\cite{oh,go2}  Most recently, Suzuki {\em et al} observed a
long-range magnetic order developing below 45 K in La$_{1.88}$Sr$_{0.12}$CuO$_{4}$.\cite{suzuki}
Recently, a stripe mechanism\cite{john1,john2,na,zhou,john3} has been proposed to explain the suppression of
the superconductivity. In the low temperature stripe phase, the doped
holes are concentrated in domain walls separating antiferromagnetic
antiphase domains. Stripes becomes pinned or immobilized only in the
LTT phase but not in the LTO phase.\\

Motivated by a successful observation of the spin-ordered phase at low temperatures in
La$_{1.67}$Sr$_{0.33}$NiO$_{4}$ by Raman scattering,\cite{blum1,yam}
we performed a Raman measurement on the magnetic scattering at low temperatures in 
La$_{2-x}$Sr$_{x}$CuO$_{4}$ crystals and found strong two-magnon scattering
from a La$_{1.9}$Sr$_{0.1}$CuO$_{4}$ crystal. To distinguish the magnetic Raman scattering in La$_{1.9}$Sr$_{0.1}$CuO$_{4}$ from that in
the parent compound La$_{2}$CuO$_{4}$ under the same laser excitation, we also performed a Raman 
measurement on La$_{2}$CuO$_{4}$.
Two-magnon Raman scattering has been intensively studied
in the parent compound La$_{2}$CuO$_{4}$.\cite{sugai1,lyn,sul,singh1}
A very broad peak around 3200 cm$^{-1}$ and a long tail at higher energy have
been reported.
Many theories have been developed to explain the broad Raman line
shape.\cite{singh1,girvin,chub1,chub2,morr,sand}
Singh {\em et al.}\cite{singh1} evaluated the first three frequency
moments and cumulants of the line shape, which provided a quantitative,
parameter-free check on the theoretical prediction for the effect of
quantum fluctuations (QF) in the system.\cite{singh1}
The Fleury-Loudon theory\cite{fleury} in the spin wave formalism
was examined again by several groups\cite{girvin,chub1,chub2,morr,sand}
Using these theories the authors obtained the superexchange constant J=1030 cm$^{-1}$,
which is in good agreement with the value obtained from neutron scattering
measurement.\cite{hayden,bou}
However the total line shape given by the theories is still in rather poor
agreement with  the experimental line shape. 
Recently, Chubukov, Frenkel and Morr\cite{chub1,chub2,morr} published a
triple resonance theory, based on the resonance Raman scattering theory
in the spin density wave formalism. The theory successfully explained the dependence of the
two-magnon peak intensity on the incoming photon frequency, which was
one of the key experimental puzzles.\cite{liu,blum2}  \\

The Raman phonon spectrum of La$_{2}$CuO$_{4}$ is also strongly dependent on the
frequency of the incoming photon. It is dominated by very strong two-phonon
scattering when the incoming photon is close to the
charge-transfer gap 2$\Delta$ (2.0 ev),\cite{liu,oha,yos}  while the two-magnon
scattering vanishes there\cite{liu,yos} and instead has a maximum when the photon
has energy equal to  2$\Delta$+3J.\cite{chub1,chub2,liu,blum2} Some
single phonon scattering peaks have been thought to originate from
Brillouin zone boundary phonons.\cite{yos,weber} 
Some Raman-forbidden phonons have also been obseved.\cite{yos,weber,sugai2}
Heyen, Kircher and Cardona\cite{heyen} careful examined the Raman
scattering from YBa$_{2}$Cu$_{3}$O$_{6}$ using different lasers. They explained
the appearance of four usually-infrared-active-only LO phonons in the Raman
spectrum as a result of a Fr\"{o}hlich-interaction. They also pointed out that
the unexpected peaks found in La$_{2}$CuO$_{4}$ were
Fr\"{o}hlich-interaction induced LO phonons.\\

In this paper, a much narrower two-magnon Raman peak around 3400 cm$^{-1}$
is observed below 37 K from single crystal
La$_{2-x}$Sr$_x$CuO$_4$ (x=0.1), which has a suppressed T$_c$ of 12 K. 
The magnetic Raman intensity
increases with decreasing temperature and persists at 2 K.
The Raman line shape is fitted within the Loudon-Fleury Raman
theory and the superexchange constant J was obtained. For La$_{2}$CuO$_{4}$,
strong single-phonon scattering from the whole Brillouin zone is reported.
Seven intense usually-infrared-active LO phonons are observed,
which may became Raman active via the Fr\"{o}hlich interaction. \\

\section{Experiment}
La$_{1.9}$Sr$_{0.1}$CuO$_{4}$ and one of the La$_{2}$CuO$_{4}$ crystals were prepared
using spontaneous crystallization from a CuO flux.\cite{schirmer}
The other La$_{2}$CuO$_{4}$ crystal was prepared at Bell laboratories.
After annealing in Ar gas, the crystal has a T$_{N}$=315 K,
close to the accepted value T$_{N}$=320 K for the stoichiometric compound.
The Raman measurements were made on the interior
surface of the crystals exposed by mechanical fracture  and repeated on
the polished surfaces from both sides of the crystals.\\

The Raman measurements were performed with a Bruker RFS 100 Fourier
Raman spectrometer, which operates with an infrared diode-pumped
Nd:YAG laser with wavelength of 1064 nm.  Most of the low temperature
measurements were taken with an Air-Products Heli-Tran refrigerator with a 
polypropylene vacuum-shroud window.
The sample was mounted on the copper disc with APIEZON grease.
Two silicon diodes for temperature measurement and control were mounted
on the cold finger, with one immediately adjacent to the sample.
The laser power was 6 mW at low temperatures. If the laser power
was raised above 60 mW, the magnetic Raman scattering was no longer observed.
The lowest temperature measurement at 2 K was performed in a Janis
refrigerator with the sample completely immersed in 
superfluid liquid helium. A back-scattering geometry was employed.
At low temperatures, the linearly polarized incident beam was focused on the $ab$ plane
of the crystal surface.  The spectrum was independent of the laser spot location.\\

\begin{figure}[tbp]
\epsfxsize=9.5cm
\centerline{\epsffile{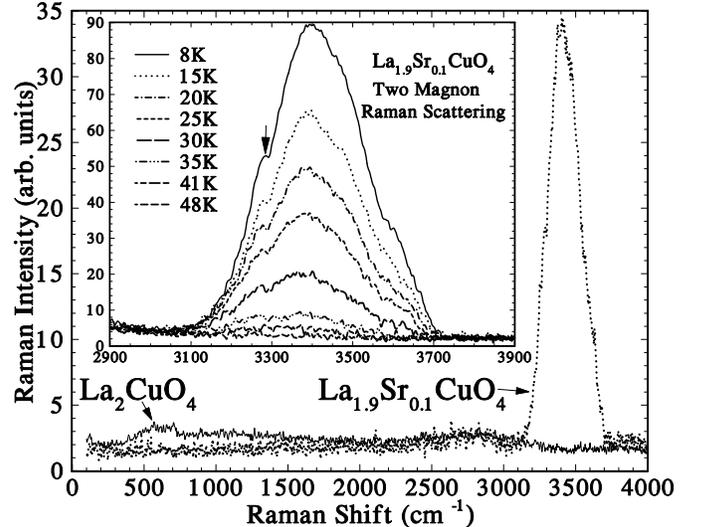}}
\caption{
Raman spectra of La$_{1.9}$Sr$_{0.1}$CuO$_{4}$ at 2 K and La$_{2}$CuO$_{4}$
at 8 K between 100 and 4000 cm$^{-1}$ with a laser power of 5 mW and an average
of 1,200 scans. The insert shows the temperature dependent two-magnon Raman
scattering in La$_{1.9}$Sr$_{0.1}$CuO$_4$ between 2900 and 3900 cm$^{-1}$.
The spectra look much narrower than the two-magnon scattering
around 3200 cm$^{-1}$ from the parent compound La$_2$CuO$_4$
under the excitation of a visible laser, and have a strong
temperature-dependence. An arrow indicates another possible peak.
}
\label{fig1}
\end{figure}

\section{Results}
Fig.\ \ref{fig1} shows the Raman spectrum of La$_{1.9}$Sr$_{0.1}$CuO$_{4}$ 
at 2 K with a very strong and sharp peak centered at 3419 cm$^{-1}$. This is
two-magnon scattering which has been intensively investigated by experiments in 
La$_2$CuO$_4$ and YBa$_2$Cu$_3$O$_{6+\delta}$\cite{sugai1,lyn,sul,singh1,blum2}
and theories.\cite{chub1,chub2} For comparison,
the spectrum of La$_{2}$CuO$_{4}$ is also showed in the figure under the
same conditions. It may be seen that the two-magnon
scattering in La$_{2}$CuO$_{4}$ is not observed with the infrared laser. This is
because the laser frequency of 9,394 cm$^{-1}$ is below the charge transfer
gap of 16,100 cm$^{-1}$ (2.0 ev),\cite{yos} and so the intensity of
two-magnon scattering will be very weak.
This dependence has been observed experimentally\cite{liu,blum2} and explained by the triple
resonance theory.\cite{chub1,chub2,morr}  The reappearance of the two-magnon scattering
in La$_{1.9}$Sr$_{0.1}$CuO$_{4}$ can be explained by the reemergence of
the magnetic order at low temperatures in La$_{2-x}$Sr$_{x}$CuO$_{4}$
with x around 0.11.\cite{wat,kuma,oh,go1,go2,suzuki}  
The different bahavior of two-magnon Raman scattering with the infrared laser
between La$_{1.9}$Sr$_{0.1}$CuO$_{4}$ and  La$_{2}$CuO$_{4}$  indicates
that the magnetic structure in La$_{1.9}$Sr$_{0.1}$CuO$_{4}$ is different 
from that in La$_{2}$CuO$_{4}$.\\

\begin{figure}[tbp]
\epsfxsize=9.5cm
\centerline{\epsffile{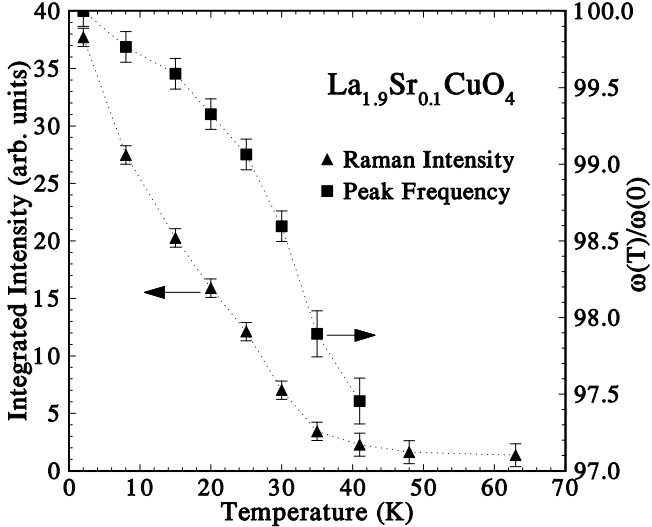}}
\caption{
Two-magnon Raman intensity and peak frequency as a function of temperature.
The integrated intensity was obtained from the averaged results both with and without
the background scattering. We obtained the peak value by fitting the spectrum to a Gaussian.
}
\label{fig2}
\end{figure}

Two-magnon Raman scattering records mainly short wavelength (large {\em k}
near the Brillouin zone boundary) magnetic excitations. The magnetic interaction in the
CuO$_2$ planes can be described by a Heisenberg Hamiltonian on
the two-dimensional square lattice:
$H=J \sum_{<i,j>}{\bf S}_i \cdot{\bf S}_j$,
where {\bf S}$_i$ is the spin-1/2 operator at site i and the summation is
over nearest-neighbor Cu pairs. The scattering process involves a
photon-stimulated virtual charge-transfer excitation that exchanges
two-spins.\cite{chub2} A value of the exchange constant J can be extracted from
the Raman scattering. Spin-wave theory\cite{girvin,chub2,sand,parkin}
predicts a peak value at $\omega_p$=2.78J$_{eff}$.
Considering the spin-wave interaction on the magnon, we must apply a
relationship J$_{eff}$=Z$_c$J with a renormalization factor Z$_c$ which
is averaged to be 1.168 from results calculated by the spin wave theory
and series expansion estimates.\cite{girvin,singh2,oguchi}  
We measured $\omega_p$=3419 cm$^{-1}$.
Thus we get J=1052 cm$^{-1}$. This is in very good
agreement with J=1042 cm$^{-1}$ (1500 K) which was used to fit the temperature dependence
of the g factor in the electron-paramagnetic resonance (EPR) measurement of La$_{1.9}$Sr$_{0.1}$CuO$_{4}$.\cite{joerg} \\

The insert of Fig.~\ \ref{fig1} shows the two-magnon Raman spectra of
La$_{1.9}$Sr$_{0.1}$CuO$_{4}$  between 2900 and 3900 cm$^{-1}$ as a
function of temperature. We note from Fig.\ \ref{fig1} the following properties
of the Raman feature: (i) The two-magnon Raman intensity has a strong temperature
dependence. However no appreciable temperature variation of the two-magnon
scattering has been observed in La$_2$CuO$_4$ for $0<T<300$ K;\cite{sugai1,singh1}
(ii) The Raman line (covering 3100 to 3700 cm$^{-1}$) looks very narrow
compared with that from La$_2$CuO$_4$ which has a broad Raman feature
extending from 2000 to over 7000 cm$^{-1}$;
(iii) The Raman peak is more symmetric than that in La$_2$CuO$_4$ as we will
show by calculating the first three cumulants.\cite{singh1} \\

\begin{figure}[tbp]
\epsfxsize=9.5cm
\centerline{\epsffile{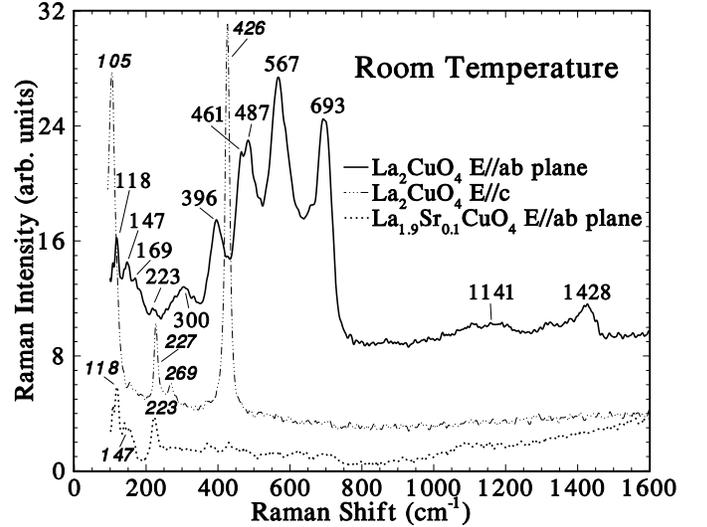}}
\caption{
Raman spectra of La$_{2}$CuO$_{4}$ and La$_{1.9}$Sr$_{0.1}$CuO$_{4}$
at room temperature between 100 and 1600 cm$^{-1}$ with a laser power
of 25 mW and an average of 12,000 scans.
}
\label{fig3}
\end{figure}

The temperature-dependent integrated Raman intensity, as well as the peak
frequency, are plotted in Fig.\ \ref{fig2}.
The intensity decreases with increasing temperature up to 37.
It is clear that the short-range
antiferromagnetic order begins to be established at about 37 K, which is
close to the temperature where the magnetic order has been
reported by NMR and neutron-scattering for x=0.11 and 0.12\cite{go2,suzuki}
in La$_{1-x}$Sr$_x$CuO$_4$
The magnetic scattering is still seen at 2 K below T$_c$, which suggests a
local coexistence of superconductivity and magnetic ordering as has been seen
in other materials.\cite{na,john3,two}  It is also shown in Fig.\ \ref{fig2}
that the peak frequency decreases as the temperature is increased.
The frequency is reduced by 2.5$\%$ at 37 K, which is very similar to
the behavior of the two-magnon peak in the two-dimensional antiferromagnet K$_{2}$NiF$_{4}$
\cite{fleury2} where the spin wave theory of two-magnon
interaction works very well.\cite{parkin} The two-dimensional system is
unlike three-dimensional compounds which have a stronger temperature
dependence of the peak frequency.\cite{fleury2,spin1}  We should point out
that at higher temperatures the La$_{1.9}$Sr$_{0.1}$CuO$_{4}$ crystals have the K$_2$NiF$_4$-type
structure.\\

Fig.\ \ref{fig3} shows the polarized room-temperature Raman spectra below 1600 cm$^{-1}$.
Two-phonon scattering in La$_{2}$CuO$_{4}$, which is the main feature
observed under the excitation of visible lasers, becomes weak here since
the infrared laser is far away from the charge-transfer gap.
The E//ab single-phonon spectrum, observed here with the
infrared laser, is unusual. Firstly, the strong peaks at 693, 567, 487,
461, and 396 cm$^{-1}$ are not expected in the Raman spectrum.\cite{yos,weber,sugai2}
Secondly, compared with the neutron scattering data, the peak at 693 cm$^{-1}$
has the value of the highest-energy phonon branch with
the k vector around (0.25, 0.25, 0),\cite{yos,chap} which is in the
middle of the Brillouin zone. It is the same for the other strong peaks
at 567, 461, and 396 cm$^{-1}$. Normally only zone-center phonons are
Raman allowed by the momentum conservation rule.
Thirdly, these strong peaks appear very broad, which probably points to
contributions from the whole Brillouin zone. Fourthly, these peak frequencies
are identical to those of infrared LO phonons as may be seen
from Table\ \ref{table1}.\\

\begin{table}[tbp]
\caption{
Frequencies (cm$^{-1}$) of the observed unexpected Raman peaks in the Raman 
spectrum compared with the calculated transverse and longitudinal optical
(TO and LO, respectively) modes from the infrared (IR) measurements.    
}
\begin{tabular}{ccccccc} 
Raman & \multicolumn{2}{c}{IR (Ref.\ \onlinecite{kovel})} 
& \multicolumn{2}{c}{IR (Ref.\ \onlinecite{gervais})} 
& \multicolumn{2}{c}{IR (Ref.\ \onlinecite{mark})}    \\
``forbidden'' & $\omega_{LO}$ &  $\omega_{TO}$ &  $\omega_{LO}$ &  $\omega_{TO}$ &
$\omega_{LO}$ &  $\omega_{TO}$ \\
\hline\hline

169  & 168 & 98  & 183 & 162 &     &     \\
300  & 260 & 168 & 250 & 220 & 300 & 148 \\
396  &     &     & 390 & 363 & 391 & 359 \\ 
461  & 460 & 229 & 463 & 320 &     &     \\
487  & 461 & 330 &     &     & 498 & 234 \\
567  & 566 & 480 & 574 & 501 & 535 & 516 \\
693  & 692 & 673 & 683 & 671 & 689 & 669 \\

\end{tabular}
\label{table1}
\end{table}

\section{Discussions}
We will discuss the following issues in this section:
(i) the magnetic order below 37 K in La$_{1.9}$Sr$_{0.1}$CuO$_{4}$.
(ii) the line shape of the two-magnon Raman scattering
and the theoretical fitting within the Loudon-Fleury theory;
(iii) the peak symmetry of the two-magnon Raman spectrum and the
quantum-fluctuation effect;
(iv) the Fr\"{o}hlich-interaction induced Raman scattering from
normally-Raman-inactive LO phonons in La$_{2}$CuO$_{4}$.\\

The charge and spin stripe phase is a common feature for nickle and manganese oxide
analogues of the copper oxides-La$_{2-x}$Sr$_{x}$NiO$_{4}$ and La$_{2-x}$Sr$_{x}$MnO$_{4}$.\cite{john4,mura}
The spin ordering is driven by the charge ordering.
Raman scattering from charge and spin ordered stripe phases has been
successfully observed in La$_{1\frac{2}{3}}$Sr$_{\frac{1}{3}}$NiO$_4$
by two research groups.\cite{blum1,yam}  The charge ordering has been observed by
electron diffraction in La$_{2-x}$Sr$_x$CuO$_4$ (x=0.115) below 104 K.\cite{ko}
In La$_{1.9}$Sr$_{0.1}$CuO$_{4}$ we observed a magnetic
order below 37 K but we did not observe the phonon changes due to the charge
ordering since the phonon scattering is very weak.
Although no direct evidence has been obtained from the Raman spectrum, it is still
possible that an ordered spin stripe phase is formed below 37 K in La$_{1.9}$Sr$_{0.1}$CuO$_{4}$.\\

The two-magnon Raman spectrum of the Heisenberg model can be discussed within traditional
Loudon-Fleury Raman theory.\cite{chub2,sand,fleury}  An effective Hamiltonian
for the interaction of light with spin degrees of freedom is given by:
\begin{equation}
 H_{LF}=\alpha\sum_{<i,j>}({\bf e}_i\cdot{\bf R}_{ij})
({\bf e}_s \cdot{\bf R}_{ij}){\bf S}_i\cdot{\bf S}_j
\label{equation1}
\end{equation}
where {\bf e}$_i$ and {\bf e}$_s$ are the polarization vectors of the incident and scattered light, $\alpha$ 
is the coupling constant, and {\bf R} is a unit vector connecting two nearest-neighbor sites i and j.\\

\begin{figure}[tbp]
\epsfxsize=9.5cm
\centerline{\epsffile{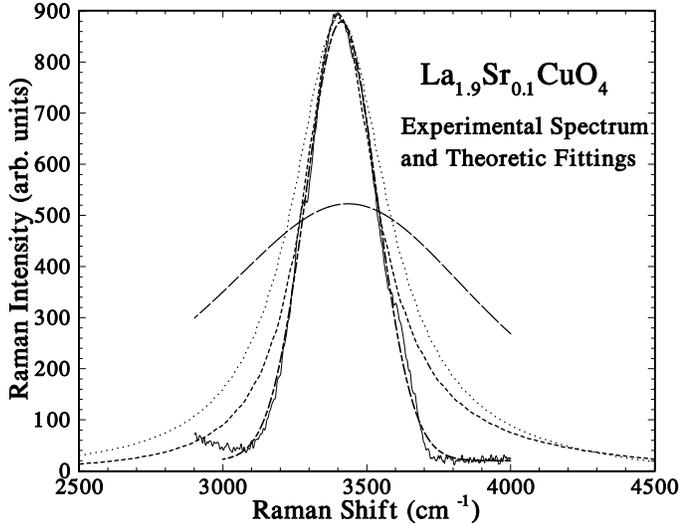}}
\caption{
The Raman spectrum and fits from various theories. The solid curve is the experimental data at 8 K. 
The long-dash curve is a fit from a formula given by Chubukov and Frenkel.\cite{chub2}  The dotted line is
a result from Canali and Girvin,\cite{girvin} which includes quantum fluctuation in the ground state. 
The short-dashed line is an exact spin-wave result by Sandvik {\em et al.}\cite{sand} 
The alt. dashed curve is a Gaussian fit.
}
\label{fig5}
\end{figure}

Chubukov and Frenkel carried out a spin-wave expansion of the profile around
its peak position by keeping S large and evaluated the results at S=1/2.
They obtained a formula for the Raman intensity for a B$_{1g}$
mode\cite{chub2} (From equation (A9) of Ref.\ \onlinecite{chub2}. 
An error in the denominator of the formula has been corrected):
\begin{equation}
R_{B_{1g}} \propto \frac { S^2\sqrt{1-\varpi^2} \ln(1-\varpi^2)}
 {\ln^2(1-\varpi^2)+\pi^2(1-(S+1)\sqrt{1-\varpi^2})^2}
\label{equation2}
\end{equation}
where $\varpi$=$\omega$/(8Z$_{c}$JS).
In Fig.\ \ref{fig5} we fitted our experimental data with the above formula
by a least-square method. We set S as a parameter
and found S=0.500024 for the best fit. The Chubukov-Frenkel formula is
successful in extracting the superexchange constant, but the line
shape given by the theory is too broad.\\

Canali and Girvin considered the quantum fluctuation effect in the
Parkinson's spin wave theory\cite{girvin,parkin} and obtained a line shape
(see Fig.\ \ref{fig4}) which is an improvement but is still a little wider
than the Raman data. Numerical
calculations of the B$_{1g}$ Raman spectrum of the two-dimensional
Heisenberg model were carried out by Sandvik {\em et al.} within the
Loudon-Fleury theory in the spin wave formalism.\cite{sand}
The exact spin wave results give a good fit to the experimental data
(see Fig.\ \ref{fig4}). Based on the above fittings, we also believe that
the main features of the spectrum are due to the Loudon-Fleury
mechanism,\cite{sand} and the quantum fluctuation effect is weak in this
system, as we will also show below in calculations of the frequency moment and
cumulant. On the other hand, we also tried to fit the peak to a
Gaussian or Lorenzian and found that Gaussians give a good fit to the peak.
Usually Lorenzians are used to fit Raman data. A good Gaussian
fitting indicates a possible multiple-peak structure of the main feature.
In Fig.\ \ref{fig2} the arrow points to a subsidiary maximum, which could
be one of many.\\

The calculations of the frequency moment and cumulant give us
information about the peak symmetry and a parameter-free check on
the quantum fluctuation effect.
The {\em n}th moment of the spectrum\cite{singh1,girvin,sand} is defined
(at T=0) as: $\rho_n=\frac{1}{I_T}\int\omega^nI(\omega)d\omega$,
where $I_T=\int I(\omega)d\omega$. The first cumulant $M_1=\rho_1$, and for $n>1$,
\begin{equation}
(M_n)^n=\frac{1}{I_T}\int(\omega-\rho_1)^nI(\omega)d\omega
\label{equation3}
\end{equation}
The first cumulant,$M_1$, gives the mean position of the Raman intensity, while the
second and third cumulants $M_2$  and $M_3$ measure the width and the asymmetry
of the line shape. Using the Raman data at 8 K,
We obtain M$_1$=3411$\pm$4 cm$^{-1}$, M$_2$=121$\pm$3 cm$^{-1}$,
and M$_3$=62$\pm$5 cm$^{-1}$,
where the uncertainties reflect the difference of calculations with and without background. 
Therefore M$_2$/M$_1$=0.035, and M$_3$/M$_1$=0.018. These values differ
considerably  from the values
( M$_2$/M$_1$=0.23, and M$_3$/M$_1$=0.26 )\cite{singh1,girvin} which were obtained including
quantum fluctuation effect, and are also three times smaller than the Parkinson's results for
S=1/2 ( M$_2$/M$_1$=0.107, and M$_3$/M$_1$=0.056). The smaller M$_2$ qnd M$_3$
values tell us that the line width is small, and that the line shape is almost
symmetrical, and therefore the quantum fluctuation effect is very weak
in La$_{1.9}$Sr$_{0.1}$CuO$_{4}$.\\

Now we try to explain the appearance of the Raman-forbidden lines in La$_{2}$CuO$_{4}$.
Motivated by the facts that the unexpected Raman lines from
YBa$_2$Cu$_3$O$_6$ were explained by Heyen, Kircher and Cardona\cite{heyen}
using a Fr\"{o}hlich interaction, and that the forbidden Raman peaks in
La$_{2}$CuO$_{4}$ have the same
values as the infrared LO phonons, we also try to assign the Fr\"{o}hlich
interaction as the cause of the additional lines in La$_{2}$CuO$_{4}$.
The forbidden first-order intraband Fr\"{o}hlich interaction
results from an expansion in the components of q of the matrix
element describing the long-range contribution to the electron-phonon
Hamiltonain.\cite{book} The Fr\"{o}hlich Hamiltonian H$_{F}$ is given by:
\begin{equation}
H_F=\frac{C_F}{|q|}(e^{i\frac{\mu}{m_e}q\cdot r}-e^{i\frac{\mu}{m_h}q\cdot r})
\label{equation4}
\end{equation}
where C$_F$ is the Fr\"{o}hlich coupling constant; m$_e$ and m$_h$ are the
effective band masses of the electron and hole, respectively, $\mu$
is the reduced mass, and r is the coordinate for the relative motion of
the electron and hole.
Upon expansion in q of the exponential in Eq.\ \ref{equation4},
the third term, proportional to $|q|$, is responsible for the
so-called forbidden LO-scattering. Because it is zero for q=0, this
term is often referred as ``forbidden". Under resonance conditions, it can
produce higher scattering efficiencies than the ``allowed" scattering, in the
case of LO-phonons in materials with inversion symmetry.\\

Several facts support the Fr\"{o}hlich model: (i) the presence of
a center of inversion in La$_2$CuO$_4$; (ii) the ``forbidden" Raman
peaks are stronger than the allowed ones and therefore greatly enhanced;
(iii) there is a strong electron-phonon interaction in La$_{2}$CuO$_{4}$ because
the phonon at 693 cm$^{-1}$, the highest frequency vibration,
has been found to have strong electron-phonon coupling;\cite{weber,weber2,sta}
(iv) the strong Raman peaks involve the scattering from Brillouin
zone-middle phonons. Thus the ``forbidden" Raman lines come from phonons
with q not equal to zero; (v) The last one is the most important fact.
As we can see from Table\ \ref{table1}, the Raman frequencies are almost the
same as the LO phonons calculated recently by Kovel {\em et al.}
in Ref.\ \onlinecite{kovel}. The Raman frequencies 169, 461, 567
and 693 cm$^{-1}$ are very close to the LO phonon energies of 168, 460,
566 and 692 cm$^{-1}$, respectively.\\

\section{Summary}
In summary, we have observed strong two-magnon Raman scattering in La$_{2-x}$Sr$_x$CuO$_4$ (x=0.1) as the
spin phase is ordered below 37 K. The line shape and the temperature dependence of the magnetic scattering 
are totally different from that observed in the parent compound La$_2$CuO$_4$. The two-magnon Raman 
intensity increases with decreasing temperatures. The temperature dependence of the peak frequency
has a two-dimension behavior. The line shape is fitted within the traditional Loudon-Fleury Raman theory
and the superexchange constant has been found to be 1052 cm$^{-1}$. No quantum fluctuation effect has been
observed in the system. The Raman scattering in La$_2$CuO$_4$ has been found to be very unusual under the
excitation of an infrared laser. We have also observed, in the Raman spectrum of La$_2$CuO$_4$,
seven usually-Raman-inactive LO phonons which possibly become Raman active via a Fr\"{o}hlich interaction. \\

\section{Acknowledgements}
The work at UBC was supported by Grant No. 5-85653 from the Natural 
Sciences and Engineering Research Council (NSERC) of Canada. \\

\end{document}